%% file: 01_main.tex
\documentclass[%
 reprint,
superscriptaddress,
 amsmath,amssymb,
 aps,
]{revtex4-2}
\usepackage{amsmath}
\usepackage{mathptmx}
\usepackage{courier}
\usepackage{newtxmath}
\usepackage{hyperref}
\usepackage{stmaryrd}
\usepackage{graphicx}
\usepackage{dcolumn}
\usepackage{bm}

\newcommand{\Cuu}{C_{u\shortleftarrow u}}
\newcommand{\Cuv}{C_{u\shortleftarrow v}}
\newcommand{\Cvu}{C_{v\shortleftarrow u}}
\newcommand{\Cvv}{C_{v\shortleftarrow v}}

\begin{document}

\preprint{APS/123-QED}

\title{Germany's current COVID-19 crisis is mainly driven by the unvaccinated}
%
%

\author{
    Benjamin F. Maier$^*$
}
    \affiliation{Institute for Theoretical Biology and Integrated Research Institute for the Life-Sciences, Humboldt-University of Berlin, Philippstr. 13, 10115 Berlin, Germany}
    \email{bfmaier@physik.hu-berlin.de}
\author{
    Marc Wiedermann
}
    \affiliation{Institute for Theoretical Biology and Integrated Research Institute for the Life-Sciences, Humboldt-University of Berlin, Philippstr. 13, 10115 Berlin, Germany}
\author{
    Angelique Burdinski
}
    \affiliation{Institute for Theoretical Biology and Integrated Research Institute for the Life-Sciences, Humboldt-University of Berlin, Philippstr. 13, 10115 Berlin, Germany}
\author{
    Pascal Klamser
}
    \affiliation{Institute for Theoretical Biology and Integrated Research Institute for the Life-Sciences, Humboldt-University of Berlin, Philippstr. 13, 10115 Berlin, Germany}
\author{
    Mirjam A. Jenny
}
    \affiliation{Harding Center for Risk Literacy, University of Potsdam, Virchowstrasse 2-4, 14482 Potsdam, Germay}
    \affiliation{Max Planck Institute for Human Development, Lentzeallee 94, 14195 Berlin, Germany}
\author{
    Cornelia Betsch
}
    \affiliation{University of Erfurt, Nordhäuserstr. 63, 99089 Erfurt, Germany}
    \affiliation{Bernhard-Nocht-Institut, Bernhard-Nocht-Straße 74,
20359 Hamburg, Germany}
\author{
    Dirk Brockmann
}
    \affiliation{Institute for Theoretical Biology and Integrated Research Institute for the Life-Sciences, Humboldt-University of Berlin, Philippstr. 13, 10115 Berlin, Germany}




\date{\today}

\begin{abstract}
Vaccines are the most powerful pharmaceutical tool to combat the
COVID-19 pandemic. While the majority (about 65\%) of the German
population were fully vaccinated, incidence started growing
exponentially in October 2021 with about 41\% of recorded new cases aged
twelve or above being symptomatic breakthrough infections, presumably
also contributing to the dynamics. At the time, it (i) remains elusive
how significant this contribution is and (ii) whether targeted
non-pharmaceutical interventions (NPIs) may stop the amplification of
the ongoing crisis. Here, we estimate that about 67\%--76\% of all new
infections are caused by unvaccinated individuals, implying that only
24\%--33\% are caused by the vaccinated. Furthermore, we estimate
38\%--51\% of new infections to be caused by unvaccinated individuals
infecting other unvaccinated individuals. In total, unvaccinated
individuals are expected to be involved in 8--9 of 10 new infections. We further show that decreasing the transmissibility of the unvaccinated by, e.~g.\ targeted NPIs, causes a steeper decrease in the effective reproduction number $\mathcal{R}$ than decreasing the transmissibility of vaccinated individuals, potentially leading to temporary epidemic control. Furthermore, reducing contacts between vaccinated and unvaccinated individuals serves to decrease $\mathcal R$ in a similar manner as increasing vaccine uptake. Taken together, our results contribute to the public discourse regarding policy changes in pandemic response and highlight the importance of combined measures, such as vaccination campaigns and contact reduction, to achieve epidemic control and preventing an overload of public health systems. 
\end{abstract}

\maketitle


\section{Introduction}

Vaccines are the most powerful pharmaceutical tool to prevent coronavirus disease 2019 (COVID-19) and combat the ongoing pandemic. Fast vaccine uptake by as many individuals as possible saves lives, people’s health, and livelihoods.
Despite large-scale vaccine roll-out campaigns, many countries, most prominently in Europe, have lately experienced a rise in case numbers and report effective reproduction numbers $\mathcal{R}$ above one for an extended period of time \cite{Whodashboard}. This means that on average, every infected person infects more than one other person, thus causing exponentially rising incidences \cite{keeling_modeling_2011}. Since the beginning of the pandemic, such resurgences have, in part, been mitigated by harsh non-pharmaceutical interventions (NPIs) such as lockdowns or curfews that limit the population’s contacts, thereby decreasing the effective reproduction number and relieving overburdened public health systems \cite{haug_ranking_2020,maier_effective_2020}. Such measures that affect large parts of the general population over a long period of time can have devastating effects, such as increasing social inequality and domestic violence, detrimental impacts on mental health, or economic disruptions \cite{habersaat_ten_2020,bajos_when_2021,mental_health,domestic_violence,ATALAN202038}. NPIs should therefore be considered a last resort of pandemic control.

Presently, many hospitals and intensive care units (ICUs) in Germany are operating at maximum capacity again or are projected to do so, soon \cite{divi}. In the four weeks between Oct 11, 2021 and Nov 7, 2021, Germany’s central public health institute, the Robert Koch Institute (RKI) reported a number of 250\,552 new symptomatic infections in individuals with known vaccination status, 90\,471 of which were attributed to vaccinated individuals, i.e. 36\% were symptomatic breakthrough cases (41\% in age groups eligible for vaccination). During this time, the average vaccination rate in age groups [0,12), [12,18), [18,60), and 60+ were 0\%, 40.1\%, 72.4\%, and 85.1\%, respectively, leading to 0\%, 4.8\%, 41.6\%, and 61.9\% of new cases being classified as symptomatic breakthrough cases within the respective age groups \cite{rki_lagebericht}, Tab.~\ref{tab:breakthrough_infections}. Simultaneously, the effective reproduction number remained at a relatively stable value of $\mathcal{R} \approx 1.2$ (under the assumption of a generation time of seven days)  \cite{RKIgithub}. 

Given that breakthrough cases are a challenge both for communication and vaccine acceptance \cite{cdc_mcmorrow} and that harsh NPIs may be illegitimate for vaccinated individuals, the current situation raises two important questions: How much does the unvaccinated population contribute to the infection dynamics despite being in the minority? And could targeted NPIs aiming at reducing the contacts of unvaccinated individuals temporarily and sufficiently suppress the infection dynamics such that harsh, large-scale NPIs could be avoided?

We address these questions based on the ``contribution matrix'' approach, a novel theoretical concept derived from the next-generation matrix approach \cite{diekmann_construction_2010}. The contribution matrix quantifies the contributions to $\mathcal{R}$ caused by the infection pathways from un-/vaccinated individuals to other un-/vaccinated individuals, considering the age and contact structure of the population, vaccination rates, as well as expected vaccine efficacies regarding susceptibility and transmission reductions, respectively. 

\begin{figure*}[t]
    \centering
    \includegraphics[width=1\textwidth]{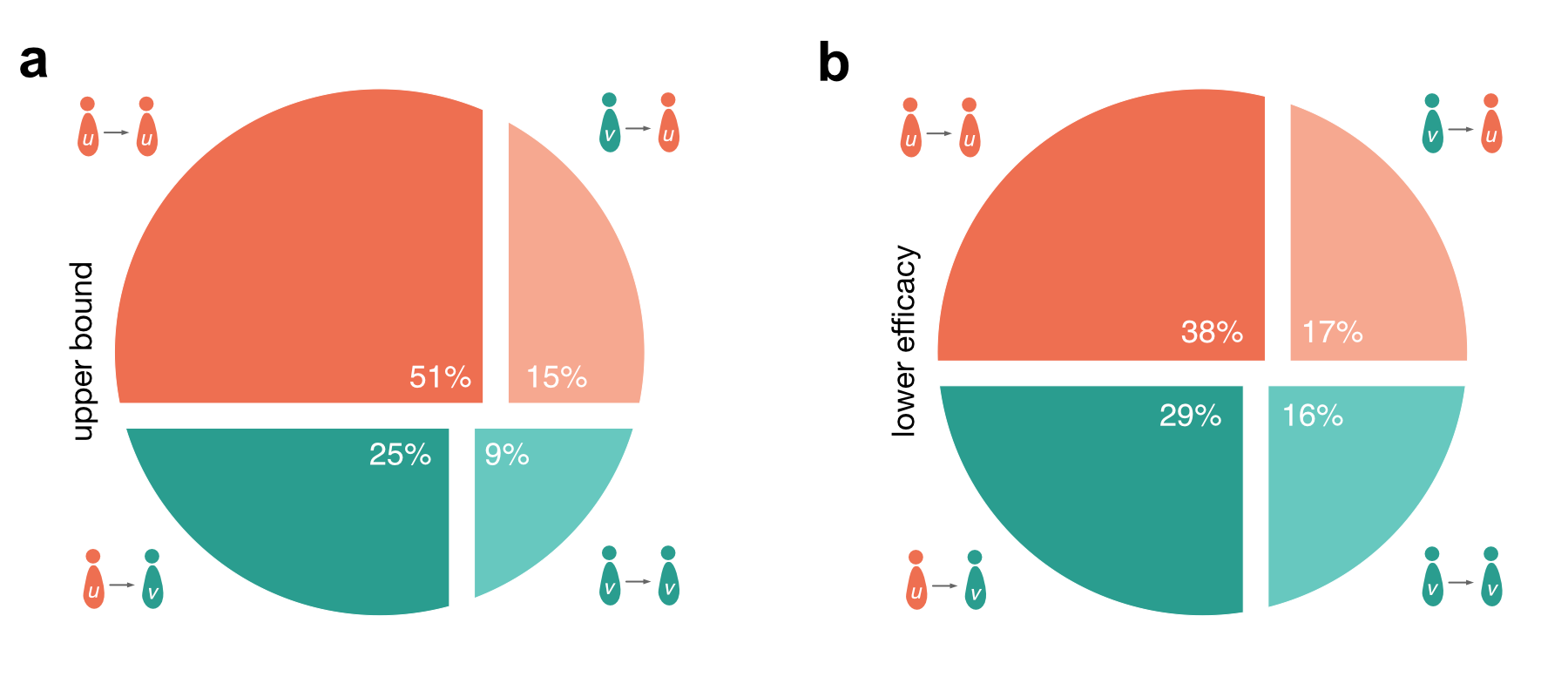}
    \caption{Estimated contributions of infection pathways to $\mathcal R$ in the (a) ``upper bound'' and (b) ``lower efficacy'' scenario as a graphical representation of Tabs.~\ref{tab:contributions-upper-bound} and \ref{tab:contributions-lower-efficacy}.
    The charts can be read as follows: Consider an infected population that caused a new generation of 100 new infecteds. Then for the left chart, 51 of those newly infected individuals will be unvaccinated people that have been infected by other unvaccinated people. Likewise, 25 newly infected individuals will be vaccinated people that have been infected by unvaccinated individuals. Hence, 76 new infections will have been caused by the unvaccinated. Along the same line, 15 newly infecteds will be unvaccinated people that have been infected by vaccinated individuals and 9 newly infecteds will be vaccinated people that have been infected by other vaccinated individuals, totaling 24 new infections that have been caused by vaccinated individuals. Considering that only 9 (a) and 16 (b) newly infecteds are vaccinated individuals that have been infected by other vaccinated individuals, we conclude that the unvaccinated are involved in 8--9 of 10 new infections as either infecting, getting infected, or both.}
    \label{fig:comprehensive-figure}
\end{figure*}

Based on this approach, we estimate that in October 2021, around 38\%--51\% of the effective reproduction number was caused by unvaccinated individuals infecting other unvaccinated individuals (see Figure~\ref{fig:comprehensive-figure}a,b). Since unvaccinated individuals have a higher probability of suffering from severe disease \cite{study_astra,study_biontech,study_moderna}, this contribution is the major factor driving the current public health crisis characterized by hospitals and ICUs reaching maximum capacity. In contrast, we estimate that only 15\%--17\% of the reproduction number can be attributed to vaccinated individuals infecting unvaccinated individuals. In October 2021, about 65\% of the population were fully vaccinated. This implies that this majority of the population contributes little to the amplification of the crisis. In total, we estimate that the vaccinated population contributes 24\%--33\% to $\mathcal{R}$ while the unvaccinated population contributes the remaining 67\%--76\%, despite the fact that unvaccinated individuals are in the minority in Germany. Considering the low number of 9\%--16\% of new infections being caused by vaccinated individuals infecting other vaccinated, we estimate that unvaccinated individuals are involved in 8--9 out of 10 new infections.

We further argue that the unvaccinated would have to reduce their transmissibility two to three times as strongly as the vaccinated in order for the system to quickly reach $\mathcal R=1$. Furthermore, decreasing mixing between individuals of distinct vaccination status can decrease $\mathcal R$. Ultimately, a higher vaccine uptake would increase the relative contribution by the vaccinated population while similarly decreasing $\mathcal R$. Combining these interventions might render the dynamics subcritical, if the vaccinated population remains at current contact behavior.

\section{Methods}

\begin{table*}[t!]
    \centering
    \begin{tabular}{c|c|c|c}
         Age group & RKI report (symptomatic cases)& Model results ``lower efficacy'' & Model results ``upper bound'' \\\hline\hline
         adolescents & 4.8\% & 21.1\% & 5.1\% \\
         adults & 41.6\% & 51.2\% & 42.3\% \\
         elderly & 61.9\% & 74.1\% & 61.5 \%
    \end{tabular}
    \caption{Share of breakthrough infections in the age groups eligible for vaccination according to official estimates and the model for ``lower efficacy'' and ``upper parameter'' scenarios.}
    \label{tab:breakthrough_infections}
\end{table*}

We use a population-structured compartmental infectious disease model that captures a variety of aspects regarding vaccination against COVID-19 to fit the data above (see SI for a full model formulation). The model’s dynamics are fully described by the next-generation matrix $K_{ji}$ of small domain, which quantifies the average number of offspring in group $j$ caused by a single infectious individual in group $i$ \cite{diekmann_construction_2010}. Here, the index $i$ (or $j$, respectively) refers to the subpopulation that is determined by a respective age group and the vaccination status within that group, thus yielding two subpopulations per age group. In the regime of small outbreaks (relative to the total population size), the ordinary differential equations governing the epidemic growth can be linearized, with the dynamics being determined by $K_{ji}$, such that the generational growth of the number of infected individuals in group $i$ follows
\begin{equation}
    y_j (g+1) = \sum_i K_{ji} y_i (g), \qquad g=0,1,2,...
\end{equation}
Both prevalence and incidence approach the eigenstate $y_i$ of $K_{ji}$ that corresponds to $\bm K$’s spectral radius, which in turn is equal to the effective reproduction number \cite{diekmann_construction_2010}. Hence, the entries of the normalized eigenvector $\hat y_i=y_i/\sum_j y_j$ contain the relative frequency of infected individuals in age/vaccination group $i$.

Consequently, the number of $j$-offspring caused by $i$-individuals in a dynamical system defined by $K_{ji}$ is given by the contribution matrix
\begin{equation}
    C_{ji} = K_{ji} \hat y_i.
\end{equation}
Summing over all matrix elements of $C_{ji}$ yields the effective reproduction number $\mathcal{R}$ (see SI). A single matrix element $C_{ji}$ can thus be considered the contribution of the $i\shortrightarrow j$ infection pathway to the reproduction number (an operational definition and derivation of the concept can be found in the SI). The normalized contribution matrix $C_{ji}$/$\mathcal{R}$ gives the relative contributions of $i\shortrightarrow j$ infections towards $\mathcal{R}$ (and consequently, towards the total number of new infections).

We derive explicit equations for the contributions of un-/vaccinated individuals in the homogeneous case, i.\ e.\ ignoring age structure (see SI, Sec.~IC). These contributions are
\begin{align}
    \label{eq:homogeneous0}
    \Cuu &= \frac{(1-v)^{2}}{1-vs}\mathcal{R}_{u}\\
    \label{eq:homogeneous1}
    \Cuv &= \frac{v(1-v)(1-s)(1-r')}{1-vs}\mathcal{R}_{v}\\
    \label{eq:homogeneous2}
    \Cvu &= \frac{v(1-v)(1-s)}{1-vs}\mathcal{R}_{u}\\
    \label{eq:homogeneous3}
    \Cvv &=\frac{v^{2}(1-s)^2(1-r')}{1-vs}\mathcal{R}_{v},
\end{align}
where $v$ is the vaccine uptake, $s$ is the susceptibility reduction after vaccination, $r'=1-(1-r)/b$ is the adjusted transmissibility reduction (i.e. it contains the relative increase of the recovery rate after a breakthrough infection $b$ and viral shedding reduction $r$), $\mathcal R_u$ is the base transmissibility of unvaccinated infecteds, and $\mathcal R_v$ is the base transmissibility of vaccinated infecteds (both of which quantify differences in behavior in the respective groups). The total effective reproduction number is given by
\begin{align}
    \mathcal R &= \Cuu + \Cvu + \Cuv + \Cvv  \nonumber\\
    &= (1-v)\mathcal R_u + v(1-s)(1-r') \mathcal R_v.
    \label{eq:total_R}
\end{align}

In the full model, we construct the next-generation matrix of small domain (see SI, Eq.~(S3)) based on the following observations, assumptions, and estimates: We structure the population into four age groups [0,12) (children), [12,18) (adolescents), [18,60) (adults), and 60+ (elderly). Contact numbers between those age groups and subpopulation sizes were constructed based on the POLYMOD (2005) data set \cite{mossong_social_2008, POLYMOD_data} using the `socialmixr' software package  \cite{funk_socialmixr_2020} (see SI). Since vaccine efficacy is, at the current time, estimated only for the status ``fully vaccinated'' in Germany without distinguishing between different vaccines, we solely distinguish between ``unvaccinated'' and ``vaccinated'' individuals in the model, regardless of the make of the received doses. Following the example of Scholz \emph{et al.} \cite{Scholz2021Einfluss}, we further assume that children and adolescents have reduced susceptibility to the virus and a reduced base transmissibility if infected, as was observed in Germany, Israel, and Greece \cite{haas_2021,israeldattner,GREECE_STUDY}. In the discussed time frame, 14.7\%, 9.4\%, 60.2\%, and 15.7\% of new cases can be attributed to the respective age groups [0,12), [12,18), [18,60), and 60+ \cite{survstat}. In order to match this distribution approximately, we calibrate the base susceptibility (i.~e.~susceptibility without vaccination) and infectiousness of our model by assuming that children are 72\% as susceptible and 63\% as infectious as adults (72\%/81\% for adolescents), which is larger than what was observed for the wild type \cite{israeldattner, GREECE_STUDY}. However, since the currently predominant B.1.617.2 variant (Delta) was generally observed to be more infectious than the wild type \cite{ong_clinical_2021}, such an increase is plausible.

In Germany, an estimated average vaccine efficacy of 72\% against symptomatic COVID-19 in adults and the elderly was found for cases reported between Oct 11, 2021 and Nov 7, 2021 \cite{rki_lagebericht}. Vaccine efficacy for adolescents has not been reported due to the respective data currently being potentially unreliable (low number of cases). Because these efficacies were computed for symptomatic cases, we use their values as an ``upper bound'' scenario regarding vaccine efficacy in our analysis, because unreported and/or asymptomatic breakthrough infections will lower the estimated efficacies. In order to obtain breakthrough infection rates in adolescents on the order of observed symptomatic breakthrough cases we assume a vaccine efficacy of $s = 92\%$ for adolescents. Despite being comparably large, this value seems justified considering that adolescents have only recently been made eligible to receive a vaccine in Germany while most adults have been vaccinated several months ago, and vaccine efficacy has been shown to decrease over the course of a few months \cite{Cohn2021.10.13.21264966,goldberg_2021}. Regarding the infectiousness of individuals suffering from breakthrough infections, viral load of vaccinated patients suffering from symptomatic COVID-19 was reported to be at the same level as of those unvaccinated \cite{Chia2021.07.28.21261295,Riemersma2021.07.31.21261387}. Another study from the UK has found decreased infectiousness in breakthrough infections \cite{REACT1_UK}. Considering these results, we assume a conservative transmission reduction of $r = 10\%$ for breakthrough infections.
In agreement with recent literature \cite{Chia2021.07.28.21261295,thompson} we further consider that the average infectious period of breakthrough infections is shorter than for unvaccinated individuals and assume a 50\% increase in recovery rate for the vaccinated, amounting to an average infectious period that is 2/3 as long as this of unvaccinated infecteds ($b=3/2$). Such an increased recovery rate can also be caused by deliberate behavior. As individuals that are not opposed to vaccination typically adhere to protection measures more consistently \cite{cosmo56}, behavioral changes following a breakthrough infection might further decrease the effective infectious period. 
Note that together with a decreased duration of infection $b=3/2$, the adjusted transmission reduction reads $r'=1-(1-r)/b=40\%$, which is lower than a 63\% reduction that was observed for household transmissions of the Delta variant between infected vaccinated and unvaccinated individuals in the Netherlands \cite{deGier2021.10.14.21264959}.

In a second, ``lower efficacy'' scenario, we consider that vaccine efficacies against infection are in the range of 50\%--60\%. This range was reported for individuals sampled randomly from the population in the United Kingdom during the second quarter of 2021 \cite{pouwels_effect_2021}. Since vaccine efficacy is expected to decrease with age \cite{collier_age-related_2021}, we assume an efficacy against infection of $s = 60\%$ for adolescents and adults as well as $s = 50\%$ for the elderly. We consider this scenario to reflect a ``low efficacy'' for two reasons: First, in the UK, a substantially larger proportion of the ChAdOx1-S vaccine (AstraZeneca) were distributed than in Germany \cite{pouwels_effect_2021,impfdashboard} which is associated with a lower base efficacy against infection \cite{pouwels_effect_2021,Thiruvengadam_cellular_2021,sheikh_sars-cov-2_2021,Bernal_2021}. In Germany, a substantial amount of used vaccines were BNT162b2 (BioNTech/Pfizer) \cite{impfdashboard}, which is associated with a higher efficacy against infection \cite{pouwels_effect_2021,sheikh_sars-cov-2_2021,Tang2021}. Second, a recent, more novel, surveillance report from the UK reports higher vaccine efficacies against infection than the study cited above \cite{uk_vaccine_surveillance}. 

To summarize the scenarios, for the ``upper bound'' the vaccination efficacy $s$ for adolescents, adults, and elderly is assumed to be $92\%,\ 72\%,\ 72\%$, and in the ``lower efficacy'' scenario $60\%,\ 60\%,\ 50\%$, respectively.

\section{Results}

\begin{figure}
    \centering
    \includegraphics[width=1\linewidth]{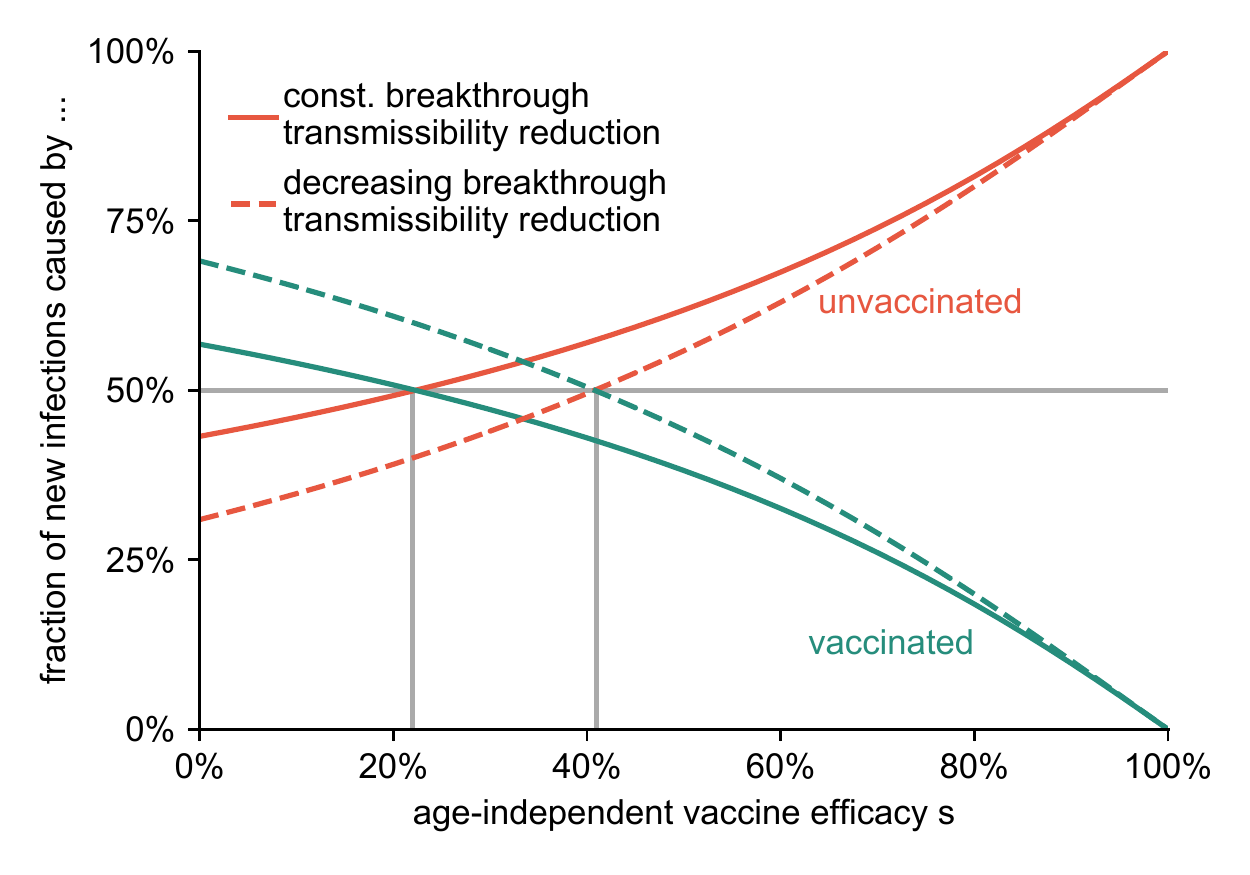}
    \caption{The fraction of new cases caused by the unvaccinated and vaccinated population for varying age-independent vaccine efficacy $s$. We consider an optimistic scenario with constant $r=0.1$ and $b=3/2$ (solid lines), and a pessimistic estimation in which $r$ and $b$ decrease according to $r=s/10$ and $b=s/2+1$ (dashed lines). As long as $s$ remains larger than approximately $22\%$ (optimistic) or $41\%$ (pessimistic), the unvaccinated minority still causes the majority of infections.}
    \label{fig:fig2_efficacy_scan}
\end{figure}

\begin{table}[t]
    \centering
    \begin{tabular}{ccc}
         & $\shortleftarrow$(u)nvaccinated & $\shortleftarrow$(v)accinated \\\hline\hline
        u$\shortleftarrow$ & 51.4\% & 15.0\% \\
        v$\shortleftarrow$ & 24.5\% & 9.1\% \\\hline
        total & 75.9\% & 24.1\%
    \end{tabular}
    \caption{Contribution to $\mathcal R$ from infections between vaccinated and unvaccinated populations for the upper parameter bounds.}
    \label{tab:contributions-upper-bound}
\end{table}

\begin{table}[t]
    \centering
    \begin{tabular}{ccc}
         & $\shortleftarrow$(u)nvaccinated & $\shortleftarrow$(v)accinated \\\hline \hline
        u$\shortleftarrow$ & 38.1\% & 17.4\% \\
        v$\shortleftarrow$ & 28.5\% & 16.0\% \\\hline
        total & 66.6\% & 33.4\%
    \end{tabular}
    \caption{Same as Tab.~\ref{tab:contributions-upper-bound} for lower vaccine efficacy.}
    \label{tab:contributions-lower-efficacy}
\end{table}

Based on the above considerations we compute the next generation matrix $K_{ji}$, the normalized population eigenvector $\hat y_i$, and the contribution matrix $C_{ji}$ for both ``lower efficacy'' and ``upper bound'' scenarios. 

As a first model validation we find that for the upper bound scenario the relative size of breakthrough infections within age groups eligible for vaccination is in good agreement with the share of current symptomatic breakthrough cases (Tab.~\ref{tab:breakthrough_infections}), albeit being slightly larger than reported values, mirroring the fact that the official number of breakthrough infections is likely affected by underreporting \cite{rki_lagebericht} and that the number of infections will be larger than the number of symptomatic breakthrough cases.

For both scenarios, we find that the dominating entry in the contribution matrix is given by the unvaccinated~$\shortrightarrow$~unvaccinated infection pathway, with a 51.4\% (upper bound) and 38.1\% (lower efficacy) contribution respectively, see Tab.~\ref{tab:contributions-upper-bound}, Tab.~\ref{tab:contributions-lower-efficacy} and Fig.~\ref{fig:comprehensive-figure}a,b. Most noteworthy, these numbers represent the largest contributions although the unvaccinated population is smaller than the vaccinated one. Moreover, the total contribution of the unvaccinated population to the effective reproduction number is 76.9\% and 64.6\% for the upper and lower efficacy scenario, respectively. In total, the unvaccinated population plays a role in 91.1\% (upper bound scenario) or 84\% (lower efficacy scenario) of cases -- either as infecting, acquiring infection, or both.

Since vaccine efficacy is expected to decrease with age and time passed
after vaccination \cite{pouwels_effect_2021}, we test how our results
change when assuming a more pessimistic susceptibility reduction of
$40\%$ for the elderly while keeping $60\%$ for all other age groups
(See SI, Sec.~IIIE). We find that our results do not change substantially, which can be attributed to the fact that the elderly generally have a lower contact behavior than other age groups.

In order to test the validity of the homogeneous approach, we further use Eqs.~(\ref{eq:homogeneous0}-\ref{eq:homogeneous3}) to compute the contribution matrix with $v=65\%$, $s=72\%$, $r=10\%$, and $b=3/2$, assuming $\mathcal R_u = \mathcal R_v$. We find relative contributions of
$\Cuu/\mathcal R=50.1\%$, 
$\Cvu/\mathcal R=26.1\%$, 
$\Cuv/\mathcal R=15.7\%$, 
$\Cvv/\mathcal R=8.1\%$, hence being in good agreement with the results of the age-structured model (cf.\ Tab.~\ref{tab:contributions-upper-bound}), showing that Eqs.~(\ref{eq:homogeneous0}-\ref{eq:homogeneous3}) can be used to estimate the order of magnitude of the contributions by the respective infection pathways. We expect this approximation to lose its validity for situations in which model assumptions become even more heterogeneous (e.g.\ strong differences in contact structure between age groups, vaccine uptakes per age group, or vaccine efficacy per age group).

\begin{figure*}
    \centering
    \includegraphics[width=1\linewidth]{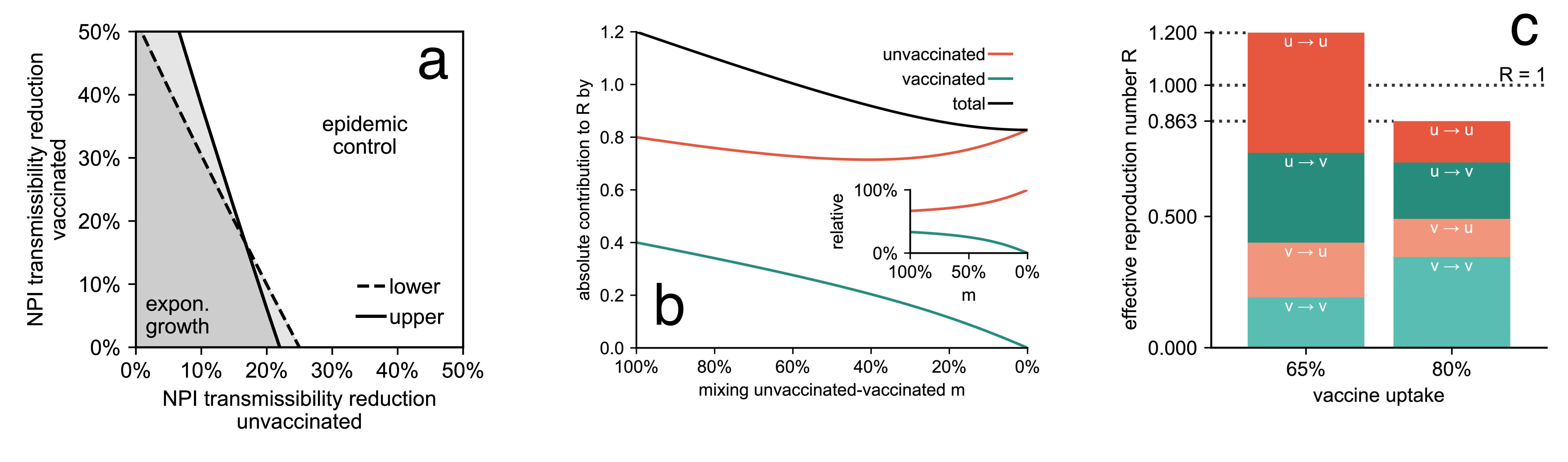}
    \caption{Efficacy of potential interventions to achieve temporary epidemic control. (a) Required additional transmissibility reduction for the unvaccinated (horizontal axis) and vaccinated (vertical axis) population to lower $\mathcal R$ to values below one, based on the assumption that the current effective reproduction number is equal to $\mathcal R=1.2$. (b) The absolute contributions to $\mathcal R$ of the unvaccinated (orange) and vaccinated population (green) as well as their sum (black) with decreasing mixing $m$ between both groups, based on the ``lower efficacy'' scenario. The inset shows the respective relative contributions. (c) Absolute contributions to $\mathcal R$ for infections between and across groups of vaccinated and unvaccinated individuals at the current vaccine uptake (left bar) and a hypothetical vaccine uptake of 80\% in the total population, i.e., 90\% in the age groups currently eligible for vaccination (right bar), based on the ``lower efficacy'' scenario. The latter would suffice to suppress $\mathcal R$ sufficiently below one, assuming that other factors determining the base transmissibility remained on the same level.}
    \label{fig:fig3_interventions}
\end{figure*}

During the period of time when vaccine efficacies were measured \cite{rki_lagebericht}, the reproduction number in Germany was reported to be at a relatively stable value of $\mathcal{R} = 1.2$ \cite{RKIgithub}. In order to achieve temporary epidemic control, it is necessary to reach a value of $\mathcal{R} < 1$ for a substantial amount of time \cite{keeling_modeling_2011}. We therefore study how the effective reproduction number would change if the transmissibility of unvaccinated individuals would be reduced. This could, for instance, be achieved by strict enforcement of contact rules regarding unvaccinated individuals at private and public gatherings that are currently already partially in place in Germany \cite{sachsen}. For our analysis we gauge $K_{ji}$ such that $\sum_{ji} (C_{ji}) = \mathcal{R} = 1.2$ for either of our two base scenarios and then individually scale the transmissibility of the vaccinated and unvaccinated to obtain those values at which the critical value $\mathcal{R} = 1$ is attained, Fig.~\ref{fig:fig3_interventions}a. We find that a transmission reduction of 22\%--25\% in the unvaccinated population would suffice to reach $\mathcal R=1$ without the need for any further restrictions. In contrast, NPIs that would affect both, vaccinated and unvaccinated to the same degree, would need to cause more than 17\% of transmissibility reduction across the entire population to achieve epidemic control. For completeness and to put numbers in perspective one may also consider the unlikely scenario where NPIs are only in place for the vaccinated population yielding a required transmissibility reduction of 52\%--73\% in that group to achieve epidemic control. The ``fastest'' way to reach $\mathcal R =1$ in the plane spanned by NPI-based transmissibility reductions in both respective subpopulations is given by the linear function that is perpendicular to the isoclines shown in Fig.~\ref{fig:fig3_interventions}a. Using the fact that the homogeneous model given by Eqs.~\eqref{eq:homogeneous0}-\eqref{eq:homogeneous3} yields acceptable approximations to the full model, we use Eq.~\eqref{eq:total_R} to derive the slope $\chi = v(1-s)(1-r')/(1-v)$ of this function (see SI, Sec.~IIIB). This quantity has to be read as ``if the unvaccinated population reduces its transmissibility by 1\%, the vaccinated population has to reduce its transmissibility by $\chi\%$ in order for the system to quickly reach $\mathcal R=1$''. With $v=65\%$, $s=72\%$, $r'=40\%$ for the ``upper bound'' scenario and $s=60\%$ for the ``lower efficacy'' scenario, we find $\chi=31\%$ and $\chi=45\%$, respectively, which suggests that in order to fairly distribute the burden of further transmissibility reductions between the respective subpopulations, unvaccinated individuals would have to reduce their current transmissibility three to two times as strongly as the vaccinated population.

We further test the robustness of our results regarding vaccine efficacy by varying an age-independent vaccine efficacy against infection that ranges from $s=100\%$ to $s=0\%$, leaving (i) $r=10\%$ and $b=3/2$ unchanged as an optimistic estimation and (ii) proportionally scaling $r=s/10$ and $b=s/2+1$ as a pessimistic estimation, while assuming vaccine uptake as reported in the Methods section. We find a monotonic decrease of breakthrough infections from non-zero values for $s=0\%$ to zero for $s=100\%$. Notably, we find that as long as vaccine efficacies do not drop below $22\%$ (optimistic) or $41\%$ (pessimistic), the majority of new cases remains to be caused by the minority of the population, which are the unvaccinated, see Fig.~\ref{fig:fig2_efficacy_scan}.

Next, we also account for the fact that the infectiousness of children and adolescents is still a matter of debate\cite{GREECE_STUDY,israeldattner,haas_2021,han_clinical_2021,cai_comparison_2020}. While for all analyses presented above we assumed reduced infectiousness for those respective age groups compared to adults and elderly, we now assume (as an upper limit) that children and adolescents are as infectious as adults. This generally leads to higher contributions by unvaccinated individuals to the overall share of infections since they represent by far the majority in these age groups. We find that the unvaccinated in this scenario cause 76.3\%--84.9\% of all new infections (see Tabs.~S1 and S2 in the SI) which is significantly larger than the 67\%-76\% obtained when susceptibility and infectiousness in children and adolescents is reduced (see again Figs.~\ref{fig:comprehensive-figure}a,b and Tabs.~\ref{tab:contributions-upper-bound},\ref{tab:contributions-lower-efficacy}. 

Moreover, we test how our results change if the assumption of homogeneous mixing between vaccinated and unvaccinated individuals is no longer met. This captures the likely scenario that vaccinated and unvaccinated populations are more prone to meet individuals of similar vaccination status rather than opposing vaccination status \cite{ernst_anthroposophy_2011,bier_simple_2015,knol}. We conceptualize this process by scaling the off-diagonal block matrices indicating offspring caused by vaccinated infecting unvaccinated individuals and vice versa with a constant factor $m\in [0,1]$ such that $m=1$ refers to our base scenario of homogeneous mixing between the two groups, Fig.~\ref{fig:fig3_interventions}a and SI Sec.~IIID. As expected we find that the relative contribution to $\mathcal R$ of the unvaccinated increases with decreasing $m$. At the same time the total value of $\mathcal R$ decreases with decreasing mixing $m$, which illustrates the efficacy of NPIs that aim at reducing contacts between individuals of distinct vaccination status.

Ultimately, we investigate how different the current situation could be if vaccine uptake was higher than 65\%. To this end, we choose the ``lower efficacy`` scenario, but increase the respective vaccine uptake for adolescents, adults, and elderly to 90\% each, leading to an 80\% uptake in the total population, Fig.~\ref{fig:fig3_interventions}c. In this case, the effective reproduction number would be lowered to a value of $\mathcal R=0.86$ instead of $\mathcal R=1.2$, implying epidemic control. Since more people would be vaccinated, both the relative and absolute contributions of vaccinated individuals to $\mathcal R$ would increase. Yet, the most important difference to the base scenario of $v=65\%$ is the reduction of the absolute contribution of unvaccinated individuals, which would decrease from $\Cuu+\Cvu=0.8$ to $\Cuu+\Cvu=0.37$ (i.e.~50\%). Because unvaccinated infecteds have a much higher probability of suffering from severe disease and being hospitalized, such a reduction can be substantial for relieving an overburdened public health system.

\section{Discussion \& Conclusion}

After vaccine rollout programs in Germany have slowed down, incidence is currently rising to unprecedented levels, with hospitals and ICUs reaching or having reached maximum capacity. As about 41\% of reported cases aged 12 or higher were recorded as breakthrough infections in October 2021, two questions naturally arise: (i) How much are the vaccinated still contributing to the infection dynamics and (ii) can targeted NPIs help achieve temporal epidemic control? 

Here, we developed a model-based framework that allows for quantifying the contributions of different infection pathways between and across vaccinated and unvaccinated groups towards the effective reproduction number $\mathcal R$. Based on this framework and reasonable assumptions regarding vaccine efficacy in line with the current literature, we conclude that about 67\%--76\% percent of the effective reproduction number are caused by unvaccinated individuals, with 38\%--51\% of its value determined by unvaccinated individuals infecting other unvaccinated individuals. We further showed that targeted NPIs that would decrease the transmissibility of unvaccinated individuals by 22\%--25\% could suppress epidemic growth reaching $\mathcal{R} < 1$, under the assumption that vaccinated individuals would continue to behave as before, i.e., with no additional NPIs in place for this respective group. Yet, it is questionable how well NPIs can be targeted towards single subpopulations, both for logistical and ethical reasons. We found that for NPIs that would affect both unvaccinated and vaccinated individuals, those that reduce the transmissibility of the unvaccinated two to three times as strongly as the vaccinated population would reduce $\mathcal R$ in the most efficient manner.

The analyses performed here represent model-based estimations that are limited by data quality and a large number of parameters that have to be estimated based on the available data. This includes epidemiological data as well as contact data from the POLYMOD study, which is by now already over 15 years old and might therefore inaccurately portray the mixing behavior of the German population. A further limiting factor is that the ascertainment of breakthrough infections might be larger than accounted for, as vaccinated infecteds experiencing mild symptoms might not be as likely to have their infection reported, thus leading to a potential overestimation of vaccine efficacy. Future empirical studies, e.g. using contact tracing data, will be necessary to confirm or refute our claims.

While we consider population mixing across age groups, we also implicitly assume homogeneous mixing between vaccinated and unvaccinated individuals in our base scenarios. Yet, the intention to vaccinate has been shown to follow rules of social contagion, rendering it likely that vaccinated and unvaccinated individuals mix less across groups. We showed that, in this case, the contribution of unvaccinated individuals to $\mathcal{R}$ would be of even larger magnitude. We want to stress that one should be careful not to overinterpret this result as explicit advice for future NPIs to increase segregation as our analysis cannot account for any psychological or socio-cultural consequences of such policies or recommendations.

Finally, an increased vaccine uptake would increase both the relative and absolute contributions that the vaccinated population makes towards $\mathcal R$ while similarly decreasing the effective reproduction number's absolute value, potentially leading to temporary epidemic control under the assumption of unchanged behavior. In light of the current slow growth of vaccine uptake in Germany \cite{impfdashboard} and low intention to vaccinate among those that are yet unvaccinated \cite{cosmo56}, such an increase in uptake, however, seems unlikely to be achieved quickly.

We furthermore stress that our results are estimations made for the comparatively short period between Oct 11, 2021 and Nov 7, 2021. As vaccine efficacy against infection has been reported to decrease with time after receiving the final dose, the contribution of the vaccinated population will thus increase in the future if no further interventions are introduced. Hence, fast and wide-spread booster vaccination is a crucial measure to avoid an increasing reproduction number and a potentially worsening public health crisis.

In summary, our results suggest that a minority of the population (i.e., the unvaccinated) contribute a substantial part to the infection dynamics, thus making them the primary driver of the currently worsening public health crisis in Germany and presumably also in other countries that are experiencing similar dynamics. We also show that this effect can be compensated through targeted NPIs that effectively lower the transmissibility of infected, yet unvaccinated, individuals. Hence, our study further underlines the importance of vaccines as a pharmaceutical intervention regarding epidemic control and highlights the importance of increasing vaccine uptake, e.g.\ through campaigning or low-threshold offers, wherever possible, in order to achieve efficient and long-term epidemic control and preventing an overload of public health systems. 

\begin{acknowledgments}
We thank L. E. Sander for valuable comments. B.F.M is financially supported by the Joachim Herz Stiftung as an \emph{Add-On Fellow for Interdisciplinary Life Science}.
\end{acknowledgments}


\input{01_main.bbl}

\pagebreak
\newpage
\ 
\newpage

\onecolumngrid
\begin{center}
  \textbf{\large Supplementary Material: Germany's current COVID-19 crisis is mainly driven by the unvaccinated}\\[.2cm]
\end{center}

\setcounter{equation}{0}
\setcounter{section}{0}
\setcounter{figure}{0}
\setcounter{table}{0}
\renewcommand{\theequation}{S\arabic{equation}}
\renewcommand{\thefigure}{S\arabic{figure}}
\renewcommand{\thetable}{S\arabic{table}}

\section{The contribution matrix}

\subsection{Contribution matrix for vaccinated populations}

Consider a generalized compartmental susceptible-infected-removed
model that also tracks vaccinated individuals and breakthrough infections.
We assume that the population can be sorted into $M\in\mathbb{N}^{+}$
groups and that a single individual can be in any of $V\in\mathbb{N}^{+}$
vaccination states. The ordinary differential equations describing
the evolution of an epidemic in this model are given as 
\begin{align*}
\partial_{t}S_{i}^{\Gamma} & =-\sum_{\Lambda=1}^{V}\sum_{j=1}^{M}\alpha_{ij}^{\Gamma\Lambda}S_{i}^{\Gamma}\frac{I_{j}^{\Lambda}}{N_{j}}\\
\partial_{t}I_{i}^{\Gamma} & =\sum_{\Lambda=1}^{V}\sum_{j=1}^{M}\alpha_{ij}^{\Gamma\Lambda}S_{i}^{\Gamma}\frac{I_{j}^{\Lambda}}{N_{j}}-\beta_{i}^{\Gamma}I_{i}^{\Gamma}\\
\partial_{t}R_{i}^{\Gamma} & =\beta_{i}^{\Gamma}I_{i}^{\Gamma}.
\end{align*}
Here, $S_{i}^{\Gamma}$ represents the compartment counting susceptibles
in population group $1\leq i\leq M$ of size $N_{i}$ that have vaccination
status $1\leq\Gamma\leq V$ (analogously, $I_{i}^{\Gamma}$ represent
the respective (breakthrough) infections, and $R_{i}^{\Gamma}$ the
respective removed individuals).

We will assume in the following that the epidemic is contained well
enough such that the fraction of infected individuals in population
group $i$ is always much lower than the respective remaining susceptibles
in this group, which means that we can linearize the equations above around the
disease-free state. We further argue that outbreaks will push the
epidemic quickly into the eigenvector corresponding to the largest
eigenvalue of the system's Jacobian. The disease-free state is, for
each population group $i$, given as
\begin{align*}
\sum_{\Gamma=1}^{\Lambda}\tilde{S}_{i}^{\Gamma} & =N_{i}\\
\tilde{I}_{i}^{\Gamma} & =\tilde{R}_{i}^{\Gamma}=0,\ \ \forall\Gamma.
\end{align*}
For simplicity, we will assume that $\Gamma=1$ refers to unvaccinated
individuals and $\Gamma>1$ refers to vaccinated individuals. For
example if we consider $V=2$ in an otherwise homogeneous population,
and assume that 60\% of a population of size $N=10^{7}$ are vaccinated,
we have the disease-free state $\tilde{S}^{\Gamma=1}=4\times10^{6}$,
$\tilde{S}^{\Gamma=2}=6\times10^{6}$ and $\tilde{I}^{\Gamma}=\tilde{R}^{\Gamma}=0.$ 

We want to derive the contributions vaccinated individuals make to
the effective reproduction number of this system in the regime of
small outbreaks. To this end, we use the next-generation-matrix framework \cite{diekmann_construction_2010,diekmann_definition_1990}. Note that we are
only interested in contributions to the reproduction number, which
remains unchanged by the introduction of latent compartments which
we can therefore safely ignore without invalidating our results.

The system's next generation matrix is derived as follows. Consider
the transmission matrix of small domain

\[
T_{ij}^{\Gamma\Lambda}=\alpha_{ij}^{\Gamma\Lambda}\tilde{S}_{i}^{\Gamma}N_{j}^{-1}
\]
and the transition matrix of small domain
\[
\Sigma_{ij}^{\Gamma\Lambda}=\beta_{i}^{\Gamma}\delta_{ij}\delta_{\Gamma\Lambda}
\]
with Kronecker's symbol
\begin{align*}
\delta_{ij} & =\begin{cases}
1 & \mathrm{if\ }i=j,\\
0 & \mathrm{otherwise.}
\end{cases}
\end{align*}
Note that both of these matrices are actually tensors of size $M\times M\times V\times V$
and capture the reproduction dynamics caused by infecteds of group
$j$ and vaccination status $\Lambda$ towards susceptibles of group
$i$ and of vaccination status $\Gamma$.

Now, the next generation matrix $\bm{K}$ is given as 
\[
\bm{K}=\bm{T}\bm{\Sigma}^{-1}.
\]
Since $\bm{\Sigma}$ is diagonal, we can write down $\bm{K}$ explicitly,
namely as 
\[
K_{ij}^{\Gamma\Lambda}=\frac{\alpha_{ij}^{\Gamma\Lambda}\tilde{S}_{i}^{\Gamma}}{\beta_{j}^{\Lambda}N_{j}}.
\]
The reproduction number of this system is given as the spectral radius
of the next generation matrix
\[
\mathcal{R}=\rho(\bm{K}).
\]
Around the disease-free equilibrium, the system can be linearized.
Consider a vector $\bm{y}$ whose elements $y_{i}^{\Gamma}$ contains
the number of infecteds in population group $i$ and vaccination status
$\Gamma$. The per-generation growth of these infecteds will effectively
follow 
\[
\bm{y}(g+1)=\bm{K}\bm{y}(g),
\]
which means $\bm{y}$ will quickly approach a state that points
in the direction of the eigenvector $\hat{\bm{y}}$ corresponding
to the largest eigenvalue $\mathcal{R}$. Let's normalize this eigenvector
such that
\[
\sum_{\Gamma=1}^{V}\sum_{i=1}^{M}\hat{y}_{i}^{\Gamma}=1.
\]
With 
\[
\bm{K}\hat{\bm{y}}=\mathcal{R}\hat{\bm{y}},
\]
this implies
\[
\sum_{\Gamma=1}^{V}\sum_{i=1}^{M}\left(\sum_{\Lambda=1}^{V}\sum_{j=1}^{M}K_{ij}^{\Gamma\Lambda}\hat{y}_{j}^{\Lambda}\right)=\mathcal{R}.
\]
From this equation, we can read off the contribution matrix
\begin{align}
C_{ij}^{\Gamma\Lambda} & =K_{ij}^{\Gamma\Lambda}\hat{y}_{j}^{\Lambda} \label{eq:contribution-matrix-full}\\
 & =\frac{\alpha_{ij}^{\Gamma\Lambda}\tilde{S}_{i}^{\Gamma}}{\beta_{j}^{\Lambda}N_{j}}\times\hat{y}_{j}^{\Lambda},\nonumber
\end{align}
which contains the contributions to $\mathcal{R}$ made by infected
individuals of group $j$ and vaccination status $\Lambda$ towards
susceptibles of group $i$ and vaccination status $\Gamma$ in the
regime of small outbreaks. To obtain average population-wide contributions
of vaccinated and unvaccinated individuals, we can simply sum over
the contributions of all population groups to find 
\begin{equation}
C^{\Gamma\Lambda}=\sum_{i=1}^{M}\sum_{j=1}^{M}C_{ij}^{\Gamma\Lambda}.
\label{eq:contribution-matrix-reduced}
\end{equation}

\subsection{Constructing the next generation matrix in presence of vaccination}

In this section we want to introduce some quantities that clarify
the construction of the infection rate matrix $\bm{\alpha}$, the
next generation matrix $\bm{K}$, and how to simplify the approach.

Let's begin with the base case, i.~e.~the absence of working vaccines
in which $V=1$ (we only have one vaccination status which is ``unvaccinated'').
We further introduce $\alpha_{j}$ as the vector quantifying population
group-specific infection rates because viral shedding could be, for
instance, age-specific, as well as the contact matrix $\gamma_{ij}$
that contains the ``typical'' number of contacts a randomly chosen
individual of group $j$ has towards individuals of group $i$. This
implies 
\[
\alpha_{ij}=\gamma_{ij}\alpha_{j}
\]
 and 
\[
K_{ij}=\gamma_{ij}\alpha_{j}\frac{N_{i}}{N_{j}}\frac{1}{\beta_{j}},
\]
i.~e.
\[
\bm{K}=\mathrm{diag}(\bm{N})\cdot\bm{\gamma}\cdot\mathrm{diag}(\bm{N})^{-1}\cdot\mathrm{diag}(\bm{\alpha})\cdot\mathrm{diag}(\bm{\beta})^{-1}.
\]
Here, $\beta_{j}^{-1}$ quantifies the average duration of the infectious
period of an individual of group $j$.

Now, let's assume that we do not know the explicit viral shedding
rate (or infectiousness), but that we can make reasonable assumptions
about their relative size (e.g.~``individuals of group $j$ are
twice as infectious as individuals of group $i$'' which would imply
$\alpha_{j}/\alpha_{i}=2$), such that 
\[
\alpha_{j}=\alpha_{0}a_{j}
\]
with $\bm{a}$ being a vector that contains these relative values.
Likewise, we can scale the relative duration of the average infectious
periods with 
\[
\beta_{j}=\beta_{0}b_{j}.
\]
What we usually do have estimates for is the basic reproduction number,
which we find as 
\begin{align*}
\mathcal{R}_{0} & =\rho(\bm{K})\\
 & =\alpha_{0}\beta_{0}\rho\left(\mathrm{diag}(\bm{N})\cdot\bm{\gamma}\cdot\mathrm{diag}(\bm{N})^{-1}\cdot\mathrm{diag}(\bm{a})\cdot\mathrm{diag}(\bm{b})^{-1}\right)\\
 & =\alpha_{0}\beta_{0}\rho\left(\bm{\gamma}\cdot\mathrm{diag}(\bm{a})\cdot\mathrm{diag}(\bm{b})^{-1}\right)\\
 & =\alpha_{0}\beta_{0}\rho_{0}
\end{align*}
with
\[
\rho_{0}=\rho\left(\bm{\gamma}\cdot\mathrm{diag}(\bm{a})\cdot\mathrm{diag}(\bm{b})^{-1}\right).
\]
Above, we used the fact that diagonal operators commute and that the
spectrum of matrices remains unchanged under base transformations.
With this result we can gauge $\alpha_{0}$ as 
\[
\alpha_{0}=\mathcal{R}_{0}\beta_{0}\rho_{0}^{-1}.
\]

Now, we introduce vaccines. Let's assume that vaccinated individuals
of vaccination status $\Lambda$ in population group $j$ have status-dependent
transmissibility reduction of 
\[
0\leq r_{j}^{\Lambda}\leq1
\]
when they suffer from a breakthrough infection. In the following we
will assume that $\Lambda=1$ corresponds to unvaccinated individuals
such that $r_{j}^{\Lambda=1}=0$. Likewise, we introduce a susceptibility
reduction for individuals of population group $i$ and vaccination
status $\Gamma$ of 
\[
0\leq s_{i}^{\Gamma}\leq1
\]
where, again, we assume that $s_{i}^{\Gamma=1}=0$. We therefore define
as the infection rate matrix 
\[
\alpha_{ij}^{\Gamma\Lambda}=\frac{\mathcal{R}_{0}\beta_{0}}{\rho_{0}}\gamma_{ij}(1-s_{i}^{\Gamma})(1-r_{j}^{\Lambda})a_{j}
\]
such that the next generation matrix is given as 
\[
K_{ij}^{\Gamma\Lambda}=\frac{\mathcal{R}_{0}}{\rho_{0}}\gamma_{ij}(1-s_{i}^{\Gamma})(1-r_{j}^{\Lambda})\frac{\tilde{S}_{i}^{\Gamma}a_{j}}{N_{j}b_{j}^{\Lambda}}.
\]
With $(b_{j}^{\Lambda})^{-1}$ we acknowledge that the average infectious
period can differ for vaccinated individuals. This result is independent
of exact values for infection rates $\alpha_{0}$ and average infectious
period $\beta_{0}^{-1}$.

We can further encapsulate restrictions imposed on (un)vaccinated
individuals by making the ``basic'' reproduction number depend on
vaccination status of infectious individuals such that 
\[
K_{ij}^{\Gamma\Lambda}=\frac{\mathcal{R}_{0}^{\mathrm{\Lambda}}}{\rho_0}\gamma_{ij}(1-s_{i}^{\Gamma})(1-r_{j}^{\Lambda})\frac{\tilde{S}_{i}^{\Gamma}a_j}{N_{j}b_j}.
\]
Regarding mixing behavior, we can introduce an unvaccinated-vaccinated mixing matrix that mimics social segregation of vaccinated and unvaccinated individuals. For simplicity, we will assume that this mixing matrix is independent of the size of the respective subpopulations and given as 
\begin{align*}
    \mu = \left(
        \begin{matrix}
            1 & m & m & \dots & m \\
            m & 1 & 1 & \dots & 1\\
            m & 1 & 1 & \dots & 1\\
            \vdots & \vdots & \vdots & \ddots & \vdots\\
            m & 1 & 1 & \dots & 1\\
        \end{matrix}
    \right)
\end{align*}
with $0<m\leq 1$. Then, the next-generation matrix is given as 
\begin{equation}
\label{eq:definition_next_generation}    
K_{ij}^{\Gamma\Lambda}=
\frac{\mathcal{R}_{0}^{\Lambda}}
{\rho_0}
\gamma_{ij}
\mu^{\Lambda\Gamma}
(1-s_{i}^{\Gamma})
(1-r_{j}^{\Lambda})
\frac{\tilde{S}_{i}^{\Gamma}a_j}{N_{j} b_j}.
\end{equation}
Hence, the parameter $m$ controls the amount of mixing between unvaccinated and vaccinated individuals with $m=1$ yielding homogeneous mixing and $m<1$ implying less mixing between unvaccinated and vaccinated states.

Note that if any $s_i^{\mathrm{unvacc}}\neq 1$ or $r_i^{\mathrm{unvacc}}\neq 1$, $\rho_0$ has to be computed as
\begin{equation*}
\rho_{0} = 
\rho\left(
    \mathrm{diag}(1-\bm{s}^0)
    \cdot
    \bm{\gamma}
    \cdot\mathrm{diag}(\bm{a})
    \cdot\mathrm{diag}(\bm{b})^{-1}
    \cdot\mathrm{diag}(1-\bm{r}^0)
\right)
\end{equation*}
where $s^0_i =s_i^{\mathrm{unvacc}}$ and $r^0_i =r_i^{\mathrm{unvacc}}$. Once $K_{ij}^{\Lambda\Gamma}$ has been constructed, we compute its spectral radius and corresponding normalized eigenvector $\hat y_j^\Gamma$ and both the full (Eq.~\eqref{eq:contribution-matrix-full}) and the reduced contribution matrix Eq.~\eqref{eq:contribution-matrix-reduced}. 

\subsection{Explicit derivation of the contribution matrix in a homogeneous population}

We devise a simplified
susceptible-infected-removed model \cite{keeling_modeling_2011}, where both (unvaccinated) susceptible ($S$) and vaccinated ($V$) individuals can be infected by the unvaccinated ($I$) and vaccinated infectious ($Y$) population. Both recover to either unvaccinated recovered, ($R$), or vaccinated recovered ($X$). 

For simplicity, we assume full immunity after recovery, which means that the duration of immunity is much longer than the time scale at which new outbreaks occur. Each variable reflects the relative frequency of the respective individuals in a population of size $N$, implying that $S+I+R+V+Y+X=1$. Accounting for potentially different infection rates $\alpha_Y$ and $\alpha_I$ of vaccinated and unvaccinated, susceptibles are depleted in a well-mixed system as 
\begin{align}
\label{eqn:dSdt_1}
\frac{dS}{dt}=-(\alpha_I I+\alpha_Y Y)S.
\end{align}
Additionally, we denote the total prevalence as $\mathcal I=I+Y$. In addition, not explicitly discriminating vaccinated and unvaccinated infected individuals yields
\begin{align}
  \label{eqn:dSdt_2}
    \frac{dS}{dt} = -\alpha_S \mathcal I S = -\beta \tilde{\mathcal{R}_{S}}\mathcal I S = -\beta \mathcal {R}_{S} \mathcal I.
\end{align}
Here, $\tilde{\mathcal{R}}_S=\alpha_{S}/\beta$ is the \textit{basic} reproduction number that represents the typical number of offspring per typical infectious individual in a fully susceptible population and $\beta$ is a generalized recovery rate. In addition, $\mathcal R_{ S}$ is the \textit {effective} reproduction number that represents the number of offspring per typical individual in the remaining susceptible population, i.e., $\mathcal{R}_S = \tilde{\mathcal{R}}_S S$. Combining Eq. \eqref{eqn:dSdt_1} and Eq. \eqref{eqn:dSdt_2} then yields a closed formula for $\mathcal R_S$ as
\begin{align}
    \mathcal{R}_{S} = \frac{(\alpha_II+\alpha_YY)S}{\beta \mathcal I} \label{eqn:R}
\end{align}
Note, that an analogous derivation can be done for the share of vaccinated individuals $V$, which yields a basic reproduction number $\mathcal R_V$ such that the total effective reproduction number reads $\mathcal R = \mathcal R_S +\mathcal R_V$.

The main goal of pandemic control is to decrease $\mathcal R$, e.g., by vaccination campaigns. Assuming a vaccine efficacy of $0\leq s\leq1$ against infection and a vaccine uptake $0\leq v\leq1$ homogeneously distributed over the whole population, the total number of unvaccinated individuals at risk of infection is given as $S=1-v$ and the total number of vaccinated individuals at risk of infection is given as $V=v(1-s)$, such that $S+V=1-vs$.

For low prevalence, i.e., $\mathcal I \ll 1$, an infectious individual will infect vaccinated and unvaccinated proportionally, such that the probability for a randomly chosen infected to have come from the unvaccinated susceptible population reads $p_I = (1-v)/(1-vs)$. Likewise, the probability for a randomly chosen infected to originate from the vaccinated population reads $p_Y=v(1-s)/(1-vs)$. We can therefore explicitly disentangle the total infectious population $\mathcal I$ into its two contributions
\begin{align}
I  =\frac{1-v}{1-vs}\mathcal{I}\quad \text{and}\quad Y  =\frac{v(1-s)}{1-vs}\mathcal{I}\label{eqn:proportions_of_infected}.
\end{align}
One principle question of this study is how infectious the vaccinated population will be towards the unvaccinated population if their respective base transmissibilities are reduced by targeted non-pharmaceutical interventions (NPIs). Hence, we set $\alpha_{Y}=\alpha_{v}(1-r)$. Here, $r$ represents the reduction of transmissibility per contact that arises from vaccination and $\alpha_v$ represents the base transmissibility of vaccinated individuals under targeted NPIs (e.g.~contact reductions). In contrast, unvaccinated individuals will transmit pathogens with a transmission rate $\alpha_{I}=\alpha_{u}$, with $\alpha_u$ representing a reduced base transmissibility caused by NPIs that target the unvaccinated population. Plugging these assumptions as well as Eq.~\eqref{eqn:proportions_of_infected} into Eq.~\eqref{eqn:R} yields
\begin{align*}
    \mathcal{R}_S &= \frac{[\alpha_u (1-v) \mathcal I+ \alpha_v(1-r)v(1-s) \mathcal I](1-v) }{\beta \mathcal I (1-vs)}\\
    &= \frac{1-v}{1-vs} [\mathcal{R}_u (1-v) + \mathcal{R}_v v(1-r)(1-s)].
\end{align*}
$\mathcal{R}$ can thus be decomposed into the respective contributions by vaccinated and unvaccinated individuals such that 
\begin{align*}
\mathcal{R}_{I\rightarrow S}&=\frac{(1-v)^{2}}{1-vs}\mathcal{R}_{u}\\
\mathcal{R}_{Y\rightarrow S}&=\frac{v(1-v)(1-s)(1-r)}{1-vs}\mathcal{R}_{v}
\end{align*}

Analogously, we find the contributions of vaccinated and unvaccinated individuals towards the infection of vaccinated individuals as
\begin{align*}
\mathcal{R}_{I\rightarrow V} & =\frac{v(1-v)(1-s)}{1-vs}\mathcal{R}_{u}\\
\mathcal{R}_{Y\rightarrow V}&=\frac{v^{2}(1-s)^2(1-r)}{1-vs}\mathcal{R}_{v}
\end{align*}
Plugging all four contributions together yields the total effective reproduction number 
\begin{align}
 \mathcal{R} & =\mathcal{R}_{I\rightarrow S}+\mathcal{R}_{Y\rightarrow S}+\mathcal{R}_{I\rightarrow V}+\mathcal{R}_{Y\rightarrow V}\nonumber \\
  & =(1-v)\mathcal{R}_{u}+v(1-s)(1-r)\mathcal{R}_{v}.\label{eq:R-eff--total}
\end{align}
The contribution matrix is given as 
\begin{align}
    \bm C = \left(
                \begin{matrix}
                    \mathcal R_{I\rightarrow S} & \mathcal R_{Y\rightarrow S}\\
                    \mathcal R_{I\rightarrow V} & \mathcal R_{Y\rightarrow V}
                \end{matrix}
    \right).
\end{align}

\subsection{Operational definition based on a two-dimensional example}

Consider $M$ coupled populations, individuals of which produce new individuals in each of these populations. A next generation matrix 
$K_{ij}$ of shape $M\times M$ contains the average offspring a single $j$ individual produces in population $i$. 

For instance, the matrix
\begin{align}
    \bm K = \left(
\begin{matrix}{}
  1  &  1 \\
  2  &  3 
\end{matrix}
\right)
\end{align}
describes a system of two populations, let's call them $A$ and $B$ with indices $i_A=0$ and $i_B=1$. In one generation (i.~e.~during its lifetime), a single $A$ individual produces, on average, $K_{00}=1$ individuals in population $A$ and $K_{10}=2$ individuals in population $B$. A single $B$ individual produces, on average, $K_{11}=3$ individuals in population $A$ and $K_{01}=1$ individuals in population $A$.

Let the vector $\bm y(g)$ of length $M=2$ contain the number of $A$- and $B$-individuals, respectively, at generation $g$. The per-generation dynamics follow
$$
\bm y(g+1) = \bm K \bm y(g).
$$
After a few generations, the system state $\bm y$ approaches the eigenvector of $\bm K$ that corresponds to its largest eigenvalue (spectral radius). We can compute the relative size of populations $A$ and $B$ as
$$
\hat {\bm y}=\left(\begin{matrix}A\\B\end{matrix}\right)=\left(\begin{matrix}0.27\\0.73\end{matrix}\right).
$$
Now, we want to define the so-called ``contribution matrix'' which quantifies the absolute contributions of each population to the reproduction of each respective population when the exponential growth (or decay) has approached the eigenstate.

Operationally, one can define the contribution matrix as follows. During a time of growth (decay), we track newborn individuals of both populations $A$ and $B$ for a few generations. Let's call the set of these individuals $\mathcal{I}$. For each individual $i\in\mathcal I$, we track the count of its offspring in the respective populations $A$ and $B$. Let's define as 
$$
\underline A(i) = \begin{cases}
                    1 & \mathrm{if\ }i\mathrm{\ belonged\ to\ }A,\\
                    0 & \mathrm{otherwise}\\
                  \end{cases}
$$
and
$$
\underline B(i) = \begin{cases}
                    1 & \mathrm{if\ }i\mathrm{\ belonged\ to\ }B,\\
                    0 & \mathrm{otherwise}\\
                  \end{cases}
$$
functions that give information about the populations individuals $i\in\mathcal I$ belonged to. Hence, $\mathcal I_A = \{i: i\in \mathcal I \wedge \underline A(i)=1\}$ and $\mathcal I_B = \{i:i \in \mathcal I \wedge \underline B(i)=1\}$ are the respective subsets of $\mathcal I$ that contain $A$ and $B$ individuals, respectively.

We further define as $\sigma_{p}(i)$ the number of $p$-offspring that individual $i$ produced during its lifetime. Then we can define the offspring matrix
$$
\bm P = \sum_{i\in\mathcal I}
    \left(\begin{matrix}
                \sigma_{A}(i)\underline A(i) & \sigma_{A}(i)\underline B(i)\\
                \sigma_{B}(i)\underline A(i) & \sigma_{B}(i)\underline B(i)
          \end{matrix}
    \right)
$$
whose entries $P_{ij}$ quantify how much $i$-offspring has been produced by $j$-individuals during the measurement period. Given the definitions of the sets above, we can also write $\bm P$ as 
$$
\bm P = 
    \left(\begin{matrix}
                \sum_{i\in I_A}\sigma_{A}(i) & \sum_{i\in I_B}\sigma_{A}(i)\\
                \sum_{i\in I_A}\sigma_{B}(i) & \sum_{i\in I_B}\sigma_{B}(i)
          \end{matrix}
    \right).
$$

The relative contribution matrix is then defined as
$$
\tilde {\bm C} = \frac{P}{\sum_{i\in\mathcal I} \big(\sigma_A(i)+\sigma_B(i)\big)}.
$$
Each entry $\tilde C_{ij}$ contains the $j$-produced number of $i$-offspring relative to the total number of offspring in the system during the measurement period.

The average number of offspring per any individual is given as 
$$
\mathcal R = \frac{1}{|\mathcal I|}\sum_{i\in\mathcal I} \Big(\sigma_A(i)+\sigma_B(i)\Big).
$$
This number is also called the ``basic reproduction number'' because it quantifies the average number of offspring per ``typical'' infectious individual. So in order to find the absolute contributions of $j$-induced $i$-offspring to the reproduction number we define the contribution matrix
$$
\bm C = R\tilde {\bm C},
$$
which evaluates to
$$
\bm C = \frac{1}{|\mathcal I|}      \left(\begin{matrix}
                \sum_{i\in I_A}\sigma_{A}(i) & \sum_{i\in I_B}\sigma_{A}(i)\\
                \sum_{i\in I_A}\sigma_{B}(i) & \sum_{i\in I_B}\sigma_{B}(i)
          \end{matrix}
    \right).
$$

We can also define the next generation matrix operationally. First, be reminded that $\mathcal I_A$ and $\mathcal I_B$ are the respective subsets of $\mathcal I$ that contain $A$ and $B$ individuals, respectively.
Then 
$$
\hat {\bm y} = \frac{1}{|\mathcal I|}\left(\begin{matrix} |\mathcal I_A|\\ |\mathcal I_B|\end{matrix}\right)
$$
describes the state of the system. To find $K$ we want to obtain the average number of $i$-offspring per active $j$ individual, i.~e.
$$
\bm K = 
    \left(\begin{matrix}
     \frac{1}{|\mathcal I_A|}\sum_{i\in I_A}\sigma_{A}(i) & 
     \frac{1}{|\mathcal I_B|}\sum_{i\in I_B}\sigma_{A}(i) \\
     \frac{1}{|\mathcal I_A|}\sum_{i\in I_A}\sigma_{B}(i) & 
     \frac{1}{|\mathcal I_B|}\sum_{i\in I_B}\sigma_{B}(i) \\
          \end{matrix}
    \right).
$$ 

We then see that
\begin{align}
\bm K\cdot\mathrm{diag}(\hat {\bm y}) &= 
     \left(\begin{matrix}
     \frac{1}{|\mathcal I|}\sum_{i\in I_A}\sigma_{A}(i) & 
     \frac{1}{|\mathcal I|}\sum_{i\in I_B}\sigma_{A}(i) \\
     \frac{1}{|\mathcal I|}\sum_{i\in I_A}\sigma_{B}(i) & 
     \frac{1}{|\mathcal I|}\sum_{i\in I_B}\sigma_{B}(i) \\
          \end{matrix}
    \right)\\
 &= \frac{1}{|\mathcal I|}\left(\begin{matrix}
     \sum_{i\in I_A}\sigma_{A}(i) & 
     \sum_{i\in I_B}\sigma_{A}(i) \\
     \sum_{i\in I_A}\sigma_{B}(i) & 
     \sum_{i\in I_B}\sigma_{B}(i) \\
          \end{matrix}
    \right)
\end{align}
so
$$
\bm K\cdot\mathrm{diag}(\hat {\bm y}) = \bm C.
$$
Note that the difference between $\bm K$ and $\bm C$ is subtle but important: While $K_{ij}$ contains the average number of $i$-offspring by a single $j$-individual, $C_{ij}$ quantifies the average number of $j$-caused $i$-offspring per individual, i.~e.~makes the important distinction to consider the relative amount of $j$-individuals in $\mathcal I$.

For our toy model, we can therefore easily quantify the contribution matrix by computing the eigenvector of the next generation matrix and plugging it into the equation above,
\[
\bm C = \left(
\begin{matrix}{}
  0.27 &  0.73\\
  0.54 &  2.20
\end{matrix}
\right).
\]
Here we see that by far the largest contribution to the reproduction number is by $B$ individuals that produce other $B$ individuals.

Note that this does not necessarily mean that, if we wanted to stifle growth altogether to induce decay, it would be enough to hinder $B$-individuals from reproducing. Let's say that we somehow manage to stop $B$-individuals from reproducing altogether, such that only $A$-individuals can produce offspring (either $A$ or $B$). This means that the next generation matrix is modified as 
\begin{align}
\bm K^* = \left(
\begin{matrix}{}
  1 & 0\\
  2 & 0
\end{matrix}
\right)
\end{align}
which means that the contribution matrix changes to
\begin{align}
\bm C^* = \left(
\begin{matrix}{}
  1/3 &  0\\
  2/3 &  0
\end{matrix}
\right)
\end{align}
with $\mathcal R=1$. So the population will stay constant over time. Note that the respective contributions by population $A$ in $\bm C^*$ are now of greater value than those in $\bm C$.

\section{Parameters and scenarios}

We report here the matrices and vectors used in the analyses in the main text, constructed based on the values and estimates reported in the Methods section.

\subsection{Population and contact data}

As described in the Methods section, we consider $M=4$ subpopulations of size
\begin{align*}
    \bm N = \left(\begin{matrix}
        9,137,232\\
        5,339,517\\
       46,495,023\\
       20,275,029
    \end{matrix} \right),
\end{align*}
with the contact matrix 
\begin{align*}
    \bm \gamma = \left(\begin{matrix}
 2.8394495 & 0.5205262 & 3.235192 & 0.6269835\\
 0.8907488 & 4.4044118 & 4.745159 & 0.4811966\\
 0.6357820 & 0.5449370 & 6.430791 & 1.0125184\\
 0.2825591 & 0.1267252 & 2.321924 & 2.1267606
    \end{matrix} \right),
\end{align*}
as constructed using the socialmixr software package \cite{funk_socialmixr_2020} based on the POLYMOD (2005) data \cite{mossong_social_2008}.

\subsection{Base epidemiological parameters}

As argued in the Methods section, we assume that children and adolescents have lower viral shedding rates if infected and set
\begin{align*}
    \bm a = \left(\begin{matrix}
                    0.63 & 0.63\\
                    0.81 & 0.81\\
                    1 & 1\\
                    1 & 1
    \end{matrix} \right).
\end{align*}
Regarding relative recovery rates, we assume that the infectious period of breakthrough infections is, on average, only 2/3 as long as the infectious period of unvaccinated infecteds, such that
\begin{align*}
    \bm b = \left(\begin{matrix}
                    1 & 1.5\\
                    1 & 1.5\\
                    1 & 1.5\\
                    1 & 1.5
    \end{matrix} \right),
\end{align*}
In consistence with the average fraction of fully vaccinated individuals, we define the disease-free state as 
\begin{align*}
    \tilde  {\bm S} = \left(\begin{matrix}
  N_1 &  0\\
  (1-0.401)N_2 & 0.401\times N_2\\
  (1-0.724)N_3 & 0.724\times N_3\\
  (1-0.851)N_4 & 0.851\times N_4
    \end{matrix} \right).
\end{align*}
we also assume homogeneous mixing between vaccinated and unvaccinated
\begin{align*}
    \bm \mu = \left(\begin{matrix}
            1 & 1 \\
            1 & 1
    \end{matrix}
    \right)
\end{align*}
and equal base transmissibility $\mathcal R^u=\mathcal R^v$.

\subsection{Scenario ``lower efficacy''}

Regarding the age-dependent susceptibility reduction we set 
\begin{align*}
    \bm s = \left(
        \begin{matrix}
            0.28 & 0.712\\
            0.28 & 0.712\\
            0.00 & 0.600\\
            0.00 & 0.500
        \end{matrix}
    \right).
\end{align*}
Note that here, a reduced base susceptibility was assumed for children and adolescents ($1-0.28=1-s=72\%$ of the value of adults). Setting a vaccine efficacy of 60\% therefore amounts to a total susceptibility reduction that is $1-(1-0.28)\times(1-0.60)=0.712$ (in relation to full susceptibility associated with adults).

\subsection{Scenario ``upper bound''}

Regarding the age-dependent susceptibility reduction we set 
\begin{align*}
    \bm s = \left(
        \begin{matrix}
            0.28 & 0.9424\\
            0.28 & 0.9424\\
            0.00 & 0.7200\\
            0.00 & 0.7200
        \end{matrix}
    \right).
\end{align*}
Note that here, a reduced base susceptibility was assumed for children and adolescents ($1-0.28=1-s=72\%$ of the value of adults). Setting a vaccine efficacy of 92\% therefore amounts to a total susceptibility reduction that is $1-(1-0.28)\times(1-0.92)=0.9424$ (in relation to full susceptibility associated with adults).

\section{Analyses} 

\subsection{Systematically decreasing vaccine efficacy}

We use Eq.~(\ref{eq:definition_next_generation}) and assume an age-independent susceptibility reduction $0\leq \sigma\leq 1$, such that
\begin{align*}
    \bm s = \left(
        \begin{matrix}
            0.28 &  \ 1-(1-0.28)\times(1-\sigma)\\
            0.28 &  \ 1-(1-0.28)\times(1-\sigma)\\
            0.00 &  \ \sigma\\
            0.00 &  \ \sigma
        \end{matrix}
    \right).
\end{align*}
In an ``optimistic'' scenario, $\bm r$ and $\bm b$ remain constant as defined above. In a ``pessimistic`` scenario, we assume that $\bm r$ and $\bm b$ are reduced proportionally to $\sigma$ with
\begin{align*}
    \bm b = \left(\begin{matrix}
                    1 & \ \sigma/2+1\\
                    1 & \ \sigma/2+1\\
                    1 & \ \sigma/2+1\\
                    1 & \ \sigma/2+1
    \end{matrix} \right),
\end{align*}
and
\begin{align*}
    \bm r = \left(\begin{matrix}
                    0 & \ 0.1\sigma\\
                    0 & \ 0.1\sigma\\
                    0 & \ 0.1\sigma\\
                    0 & \ 0.1\sigma
    \end{matrix} \right).
\end{align*}

\subsection{Decreasing the transmissibility of unvaccinated and unvaccinated with targeted NPIs}

In order to obtain Fig.~3a in the main text, we use the ``lower efficacy'' and ``upper bound'' scenario parameters and gauge $\mathcal R^u_0=\mathcal R^v_0$, such that the spectral radius of $K_{ij}^{\Lambda\Gamma}$ is equal to $\mathcal R=1.2$. Then, we linearly scale $\mathcal R^u$ from $\zeta_u=1-\mathcal R_u/\mathcal R^u_0=0 $ to $\zeta_u = 1-\mathcal R_u/\mathcal R^u_0=1/2$, numerically finding the value $\zeta_v =  1-\mathcal R_v/\mathcal R^v_0$ at which the spectral radius becomes $\mathcal R = 1$. 

The resulting isoclines marking $\mathcal R = 1$ are linear functions. In the homogeneous cases, we can use Eq.~\eqref{eq:R-eff--total} to find the parametric equation
\begin{align*}
    A = (1-v) (1-\zeta_u) + v(1-s)(1-r)(1-\zeta_v)
\end{align*}
where $A$ is a constant. We rewrite the equation above as
\begin{align*}
    \tilde A = (1-v)\zeta_u + v(1-s)(1-r)\zeta_v,
\end{align*}
giving
\begin{align*}
    \zeta_v = \tilde A' - \frac{1-v}{v(1-s)(1-r)}\zeta_u
\end{align*}
which determines the isoclines. The linear function that runs perpendicular to this function has slope
\begin{align*}
    \chi = \frac{v(1-s)(1-r)}{1-v}
\end{align*}
and is the ``fastest'' way to reach any isocline in the plane from any point in the plane.

\subsection{Assuming children are as susceptible and infectious as adults}

Here, we assume 
\begin{align*}
    \bm a = \left(\begin{matrix}
                    1 & 1\\
                    1 & 1\\
                    1 & 1\\
                    1 & 1
    \end{matrix} \right),
\end{align*}
as well as
\begin{align*}
    \bm s = \left(
        \begin{matrix}
            0.00 & 0.92\\
            0.00 & 0.92\\
            0.00 & 0.72\\
            0.00 & 0.72
        \end{matrix}
    \right)
\end{align*}
for the ``upper bound'' and 
\begin{align*}
    \bm s = \left(
        \begin{matrix}
            0.00 & 0.60\\
            0.00 & 0.60\\
            0.00 & 0.60\\
            0.00 & 0.50
        \end{matrix}
    \right)
\end{align*}
for the ``lower efficacy'' scenario, respectively. The results are presented in Tab.~\ref{tab:contributions-lower-efficacy-children-analysis} and Tab.~\ref{tab:contributions-upper-bound-children-analysis}.

\begin{table}[t]
    \centering
    \begin{tabular}{ccc}
         &$\leftarrow$(u)nvaccinated & $\leftarrow$(v)accinated \\\hline\hline
        u$\leftarrow$ & 64.3\% & 9.8\% \\
        v$\leftarrow$ & 20.6\% & 5.3\% \\
        \hline
        total & 84.9\% & 15.1\%
    \end{tabular}
    \caption{Contribution to $\mathcal R$ from infections between vaccinated and unvaccinated groups for the upper parameter bounds, considering that children and adolescents are both as susceptible and as infectious as adults.}
    \label{tab:contributions-upper-bound-children-analysis}
\end{table}

\begin{table}[t]
    \centering
    \begin{tabular}{ccc}
         & $\leftarrow$(u)nvaccinated & $\leftarrow$(v)accinated \\\hline \hline
        u$\leftarrow$ & 50.0\% & 13.2\% \\
        v$\leftarrow$ & 26.3\% & 10.5\% \\
        \hline
        total & 76.3\% & 23.7\%
    \end{tabular}
    \caption{Same as Tab.~\ref{tab:contributions-upper-bound-children-analysis} for lower vaccine efficacy.}
    \label{tab:contributions-lower-efficacy-children-analysis}
\end{table}

\subsection{Decreasing mixing between vaccinated and unvaccinated}
\label{sec:decreasing-mixing}
One may assume that the intention to vaccinate follows the rules of social contagion, such that it is likely that vaccinated and unvaccinated individuals meet each other less often than they meet individuals with whom they share their respective vaccination status. We can simulate such a hypothetical scenario using Eq.~(\ref{eq:definition_next_generation}) with
\begin{align*}
    \bm \mu = \left(\begin{matrix}
            1 & m \\
            m & 1
    \end{matrix}
    \right).
\end{align*}
We find that decreasing mixing (decreasing $m$ from $m=1$ to lower values $0\leq m<1$) between vaccinated and unvaccinated individuals decreases $\mathcal R$, but increases the relative contributions unvaccinated individuals make towards it (see Fig.~3b in the main text).

\subsection{Assuming lower vaccine efficacy for the elderly}

We base the following scenario on the ``lower efficacy'' scenario and additionally assume that the elderly have lower protection against infection by setting
\begin{align*}
    \bm s = \left(
        \begin{matrix}
            0.00 & 0.60\\
            0.00 & 0.60\\
            0.00 & 0.60\\
            0.00 & 0.40
        \end{matrix}
    \right),
\end{align*}
\begin{align*}
    \bm r = \left(
        \begin{matrix}
            0.00 & 0.10\\
            0.00 & 0.10\\
            0.00 & 0.10\\
            0.00 & 0.08
        \end{matrix}
    \right),
\end{align*}
and
\begin{align*}
    \bm b = \left(
        \begin{matrix}
            1 & 1.5\\
            1 & 1.5\\
            1 & 1.5\\
            1 & 1.4
        \end{matrix}
    \right),
\end{align*}
effectively reducing all relevant quantities regarding vaccine efficacies in the elderly by 20\% (relative to the base value). The results are shown in Tab.~\ref{tab:contributions-lower-eff-elderly}. Comparing these results with those obtained in the ``lower efficacy'' scenario (Tab.~III in the main text) shows that this lower efficacy does not change the original results substantially.

\begin{table}[t]
    \centering
    \begin{tabular}{ccc}
         &$\leftarrow$(u)nvaccinated & $\leftarrow$(v)accinated \\\hline\hline
        u$\leftarrow$ & 36.8\% & 17.4\% \\
        v$\leftarrow$ & 28.6\% & 17.2\% \\
        \hline
        total & 65.4\% & 34.6\%
    \end{tabular}
    \caption{Contribution to $\mathcal R$ from infections between vaccinated and unvaccinated groups for the lower efficacy scenario, additionally assuming that the elderly have lower protection than initially assumed.}
    \label{tab:contributions-lower-eff-elderly}
\end{table}

\subsection{Increasing vaccine uptake}

We use the ``lower efficacy'' scenario and replace the disease-free state with 
\begin{align*}
    \tilde  {\bm S} = \left(\begin{matrix}
  N_1 &  0\\
  (1-0.9)N_2 & 0.9\times N_2\\
  (1-0.9)N_3 & 0.9\times N_3\\
  (1-0.9)N_4 & 0.9\times N_4
    \end{matrix} \right).
\end{align*}
The absolute contributions that are shown in Fig.~3c of the main text are reported in Tabs.~\ref{tab:contributions-lower-efficacy-lower-uptake},\ref{tab:contributions-lower-efficacy-higher-uptake}.

\begin{table}[t]
    \centering
    \begin{tabular}{ccc}
         &$\leftarrow$(u)nvaccinated & $\leftarrow$(v)accinated \\\hline\hline
        u$\leftarrow$ & 0.458 & 0.208 \\
        v$\leftarrow$ & 0.342 & 0.192 \\
        \hline
        total & 0.8 & 0.4
    \end{tabular}
    \caption{Absolute contributions to $\mathcal R$ from infections between vaccinated and unvaccinated groups for the ``lower efficacy'' scenario.}
    \label{tab:contributions-lower-efficacy-lower-uptake}
\end{table}

\begin{table}[t]
    \centering
    \begin{tabular}{ccc}
         & $\leftarrow$(u)nvaccinated & $\leftarrow$(v)accinated \\\hline \hline
        u$\leftarrow$ & 0.158 & 0.145 \\
        v$\leftarrow$ & 0.215 & 0.345 \\
        \hline
        total & 0.373 & 0.49
    \end{tabular}
    \caption{Absolute contributions to $\mathcal R$ from infections between vaccinated and unvaccinated groups for the ``lower efficacy'' scenario, considering that vaccine uptake is 90\% for adolescents, adults, and the elderly, amounting to a total vaccine uptake of 80\%.}
    \label{tab:contributions-lower-efficacy-higher-uptake}
\end{table}

\subsection{Scenario ``very low efficacy''}

In order to test how our results would change if the population-wide vaccine efficacy against infection was even lower than in the ``low efficacy'' scenario, we set a susceptibility reduction of 40\% for the elderly and 50\% for the rest of the population, such that
\begin{align*}
    \bm s = \left(
        \begin{matrix}
            0.28 & 0.64\\
            0.28 & 0.64\\
            0.00 & 0.50\\
            0.00 & 0.40
        \end{matrix}
    \right).
\end{align*}
Note that here, a reduced base susceptibility was assumed for children and adolescents ($1-0.28=1-s=72\%$ of the value of adults). Setting a vaccine efficacy of 50\% therefore amounts to a total susceptibility reduction that is $1-(1-0.28)\times(1-0.50)=0.64$ (in relation to full susceptibility associated with adults).

We find the relative contributions given in Tab.~\ref{tab:contributions-even-lower-efficacy}. In this scenario, the unvaccinated cause 61.1\% of new infections, and the unvaccinated $\rightarrow$ unvaccinated infection pathway is still the dominant contribution (31.6\%). In total, unvaccinated individuals are involved in 79.3\% of all new infections.

\begin{table}[t]
    \centering
    \begin{tabular}{ccc}
         & $\leftarrow$(u)nvaccinated & $\leftarrow$(v)accinated \\\hline \hline
        u$\leftarrow$ & 31.6\% & 18.2\% \\
        v$\leftarrow$ & 29.5\% & 20.7\% \\
        \hline
        total & 61.1\% & 38.9\%
    \end{tabular}
    \caption{Relative contributions to $\mathcal R$ from infections between vaccinated and unvaccinated groups for the ``very low efficacy'' scenario.}
    \label{tab:contributions-even-lower-efficacy}
\end{table}

%
%
%
%
%
%

\end{document}

%% file: 01_main.bbl
%